# Investigation of Iterative Techniques for Synthesizing of the Array Factor of the Antenna Arrays

Atefe Akbari-Bardaskan
Department of Electrical Engineering, Ferdowsi University of Mashhad, Mashhad, Iran

*Abstract*—In this essay, several iterative techniques are investigated for synthesizing of the array factor of the antenna array. These iterative techniques include Richardson method, Jacobi method, Gauss-Seidel method, Successive-over relaxation (SOR) method. It is shown that these iterative techniques can be successfully applied to synthesize of the array factor of any arbitrary linear, planar, or even ring arrays. Traditionally, least square method (LSM) is used for this purpose. However, LSM cannot be directly applied for any practical arrays. In other words, LSM is not a general approach, and it should be mixed with other techniques to find the magnitude and phase of the excitation coefficients of each element of the under-studying arrays. Obtained results show that the proposed method, offers considerable improvements for this target.

## I. Introduction

Single antenna is used for several targets in engineering fields [1-6]. However, single antenna is suffer from low gain. To overcome this problem, linear and planar arrays are used. Antenna arrays are the important parts of the communication systems. Phased-array antennas are widely seen in 5G/6G systems, tracking applications, EMC/EMI devices [7-11].

This is due to the high directivity and steerable pattern of therm. However, implementation of an array is not simple and there are several challenges, such as complex beamforming networks, finding the proper excitation coefficients of the array elements, and reducing the side lobe level (SLL) of the pattern of an array [12-15]. To reduce the complexity of the beamforming network, various structures can be employed. These structures are established based on the substrate integrated waveguide (SIW), microstrip lines on artificial perforated substrate, coplanar waveguide, and non-uniform transmission lines [16-23].

The pattern of a linear or planar array is formed by selecting the proper values of the magnetic and phase of the excitation coefficients of array elements. Hence, the array factor synthesis is an important challenge in antenna engineering. So far, various analytical, iterative, and algorithm-based approaches are introduced for this purpose. Even, some of them can reduce the number of array elements [24-29].

In [30], a combination of static and dynamic successive optimization algorithm is proposed for this purpose. In this work, by introducing the shrinkage coefficient and crossover operator, it is tried to enhance the performance of the procedure. A new method established based on the Zernike polynomials is proposed to design a large planar array [31].

In other words, thie polynomial is considered as a global basis function for the phases of each elements. By this way, the number of optimization variables is extensively decreased in comparison to the number of array components. In [32], a low side lobe beam shaping technique is suggested. In this work, the array errors is regarded. So, the statistical mean approach is employed to determine a robust control of the covariance matrix, and to decrease the impact of the uncertainty characteristic of the array errors on the side lobe level region.

Nowadays, the time-modulated arrays (TMA) are widely regarded in communication systems. Time-modulated array can improve the flexibility of the array. To this end, time switches are employed. But, it should be noted that this type of arrays are more sensitive with respect to the unplanned errors due to the using of several time switches. In [33], an anti-error robust pattern synthesis technique is proposed. This algorithm is established based on the convex optimization method. In this work, it is shown that by the proposed algorithm, the average value of the SLL, and also the final dynamic range are extremely reduced.

In this essay, first, the problem is converted into a system of linear equations. Then, the classical least square method is defined for the problem. In the following, it is shown that several iterative techniques, including Richardson, Jacobi, Gauss-Seidel, Successive-over relaxation (SOR) method can be effectively used to solved the obtained system of equations to achieve the accuracy of the final solution. To evaluate the performance of the proposed method, several examination are considered. The obtained results confirm that these iterative techniques provide a good accuracy for the under-studying problem.

## II. Mathematical Modeling

As shown in figure (1), first, a linear, equally-spaced array is regraded. This array is oriented along z-axis. It is assumed that the number of elements is $N$, and the elements spacing is $d$. So, the array factor ($F$) is described as the following equation [34].

$$F(z) = \sum_{n=1}^{N} I_n z^{n-1} \qquad (1)$$

where,

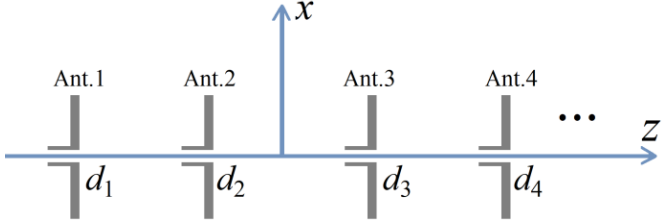

Fig. 1: The geometry of a linear array.

$$z = \exp j(kdu + \beta) \quad (2)$$

$$u = \cos\theta \quad (3)$$

In the above equations $k$ is the wave number, $\theta$ is the elevation angle in spherical coordinate, $\beta$ is the phase difference, and $I_n$'s; $n=1, 2, \ldots, N$, are the array weights. By sampling the array factor in uniform step, a linear system of equations is defined as follows.

$$\mathbf{Z}_{M \times N} \mathbf{I}_{N \times 1} = \mathbf{F}_{M \times 1} \quad (4)$$

where,

$$\mathbf{Z}_{M \times N} = [Z_{mn}]_{M \times N} = \begin{bmatrix} 1 & z_1 & \cdots & z_1^{N-1} \\ 1 & z_2 & & z_2^{N-1} \\ \vdots & & \ddots & \vdots \\ 1 & z_M & \cdots & z_M^{N-1} \end{bmatrix}_{M \times N} \quad (5)$$

$$\mathbf{I}_{N \times 1} = [I_n]_{N \times 1} = [I_1 \quad I_2 \quad \cdots \quad I_N]^T_{N \times 1} \quad (6)$$

$$\mathbf{F}_{M \times 1} = [F(z_1) \quad F(z_2) \quad \cdots \quad F(z_M)]^T_{M \times 1} \quad (7)$$

The system is linear, because each variable is in the first power only. According to Nyquist theorem, the value of $M$ is determined as follows [27].

$$M \simeq \frac{4(N-1)d}{\lambda} \quad (8)$$

where $\lambda$ is the wavelength. Target is minimizing the norm vector as.

$$\mathbf{e} = \mathbf{ZI} - \mathbf{F} \quad (9)$$

So, the problem is to minimize the objective function as follows.

$$J = \frac{1}{2}\|\mathbf{e}\|^2 = \frac{1}{2}\|\mathbf{ZI} - \mathbf{F}\|^2 \quad (10)$$

Then, solution is determined by setting the derivative of this objective function with respect to $\mathbf{Z}$ to zero.

$$\frac{\partial}{\partial \mathbf{Z}} J = \mathbf{Z}^T [\mathbf{ZI} - \mathbf{F}] = 0 \quad (11)$$

The above equations leads to the following one [34].

$$\mathbf{I} = (\mathbf{Z}^T\mathbf{Z})^{-1}\mathbf{Z}^T\mathbf{F} \quad (12)$$

In equation (4), each equation has the same value. This leads to itself biasing problem. So, equation (12) is self-biased. To overcome this challenge, iterative techniques can be employed. To this end, first, the real and imaginary parts of the problem is separated as follows.

$$\mathbf{G}_{(2m-1) \times n} = \text{real}(Z_{m \times n}) \quad (13a)$$

$$\mathbf{G}_{2m \times n} = \text{imag}(\mathbf{A}_{m \times n}) \quad (13b)$$

$$\mathbf{C}_{2m-1} = \text{real}(\mathbf{B}_m) \quad (13c)$$

$$\mathbf{C}_{2m} = \text{imag}(\mathbf{B}_m) \quad (13d)$$

As a result, the new system of linear equations is acquired.

$$\mathbf{GI} = \mathbf{C} \quad (14)$$

Now, we can use QR decomposition as follows.

$$\mathbf{G} = \mathbf{QR} \quad (15)$$

$$\mathbf{RI} = \mathbf{Q}^T\mathbf{C} \quad (16)$$

Now, several iterative techniques can be applied. It should be noted that the coefficient matrix $\mathbf{Z}$ is partitioned as follows [35]:

$$\mathbf{Z} = \mathbf{D} - \mathbf{L} - \mathbf{U} \quad (17)$$

$$\mathbf{D} = \text{diag}(\mathbf{Z}) \quad (18)$$

Also, $\mathbf{L}$, and $\mathbf{U}$ are the negative of the strictly lower and upper triangular part of $\mathbf{Z}$, respectively. The final solution of the problem using the Jacobi, Richardson, Gauss-Seidel, and Successive-over relaxation (SOR) technique are as follows.

$$\mathbf{DI}^{t+1} = (\mathbf{L} + \mathbf{U})\mathbf{I}^t + \mathbf{F}; \qquad Jacobi \quad (19)$$

$$\mathbf{I}^{t+1} = (\mathbf{1} - \mathbf{Z})\mathbf{I}^t + \mathbf{F}; \qquad Richardson \quad (20)$$

$$(\mathbf{D} - \mathbf{L})\mathbf{I}^{t+1} = \mathbf{UI}^t + \mathbf{F}; \qquad Gauss-Seidel \quad (21)$$

$$(\mathbf{D} - \omega\mathbf{L})\mathbf{I}^{t+1} = \omega(\mathbf{UI}^t + \mathbf{F}) + (1-\omega)\mathbf{DI}^t; \quad SOR \quad (22)$$

where $\omega$, $T$ are the coefficient factor and total number of iterations, respectively.

The proposed technique is straightforwardly expanded to an equally-spaced, planar array. It is assumed that the number of elements and distance between two elements in x and y direction are $N$ and $M$, and $d_x$ and $d_y$, respectively. The array factor of a planar array, placed in x-y plane, is as follows.

$$F_{planar} = \left(\sum_{n=1}^{N} I_n z_x^{n-1}\right)\left(\sum_{n=1}^{M} I_m z_y^{m-1}\right) \quad (23)$$

$$z_x = \exp j(kd_x u_x + \beta_x) \quad (24a)$$

$$z_y = \exp j(kd_y u_y + \beta_y) \quad (24b)$$

$$u_x = \sin\theta \cos\varphi \quad (24c)$$

$$u_y = \sin\theta \sin\varphi \quad (24d)$$

$$F_{planar} = \sum_{n=1}^{N}\sum_{m=1}^{M} I_{nm} z_x^{n-1} z_y^{m-1} \quad (25)$$

where

$$I_{nm} = I_n I_m, \quad n=1,...,N, \quad m=1,...,M \quad (26)$$

where $\varphi$ shows the azimuth angle in spherical coordinate. The studies show that the accuracy of all iterative methods are the same. So, only one of them are reported in the next section.

### III. NUMERICAL EXAMPLES AND DISCUSSION

Now, different and practical arrays are examined to verify the effectiveness of the method. For each example, a MATLAB program is written for implementing of the method.

#### A. Sum pattern

An equi-ripple array factor with $N=20$, $SLL=-25$dB and $d=\lambda/2$ is regarded as the first case. The acquired results, including the target array factor, synthesized array factor, and array weights are shown in Figures (1) and (2), respectively. It is seen that the array weights are not complex. Also, it can be seen that the accuracy of is good.

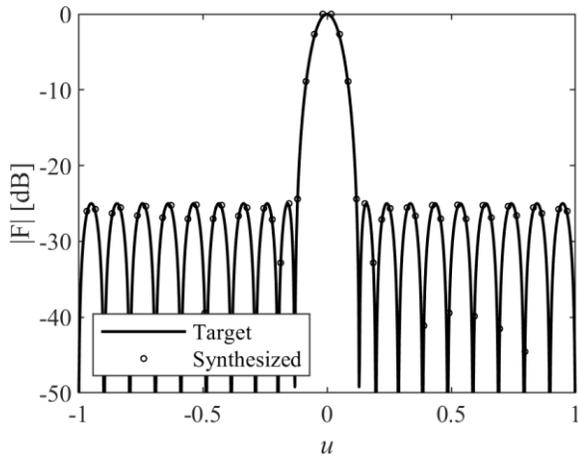

Fig. 1: The synthesized result of array with Dolph-Tschebyscheff pattern.

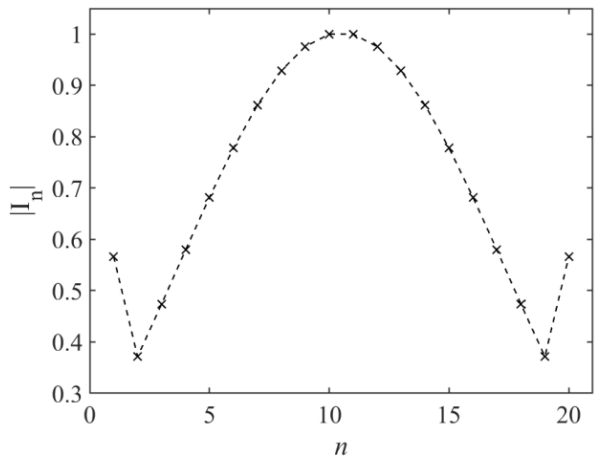

Fig. 2: The excitation coefficients of the first example.

#### B. Difference pattern

An array with difference pattern with the parameters $SLL=-30$dB, $n_p=5$, $N=15$ and $d=\lambda/2$ is tested as the second case. Figure (3) shows the target and synthesized array factors. It is seen that the obtained results are matched completely. It is well-known that an array with difference pattern has the complex excitation coefficients. The magnitude and phase of the array weights are displayed in Figure (4).

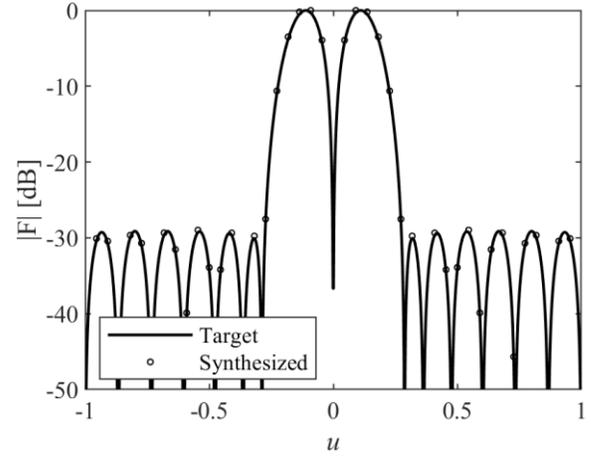

Fig. 3: The synthesized result of array with difference pattern.

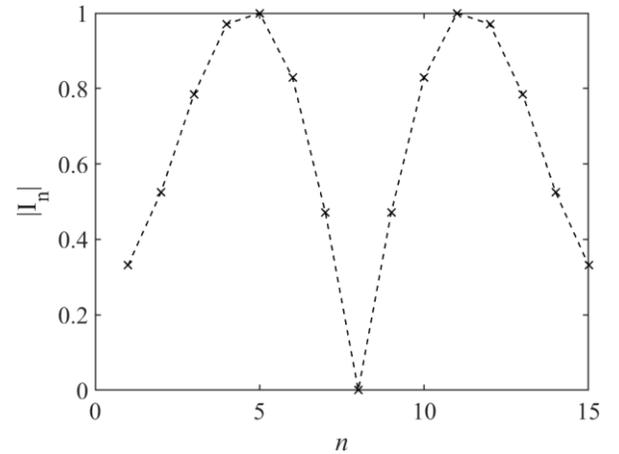

Fig. 4a: The excitation coefficients of the second example.

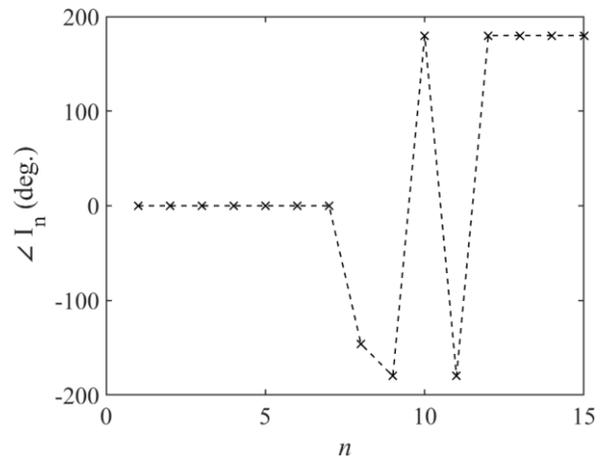

Fig. 4b: The excitation coefficients of the second example.

## C. Pattern with deep null

A linear array with $N=21$, $d=\lambda/2$, $SLL_{min}=-40$dB, and $SLL_{max}=-16$dB is considered as the third case. It is desired to synthesize of array factor with two different SLLs (two deep nulls). Target and synthesized array factors are reported and compared in Figure (5). It is seen that two array factors are matched very well. Also, the computed array weights are shown in Figure (6), which are real.

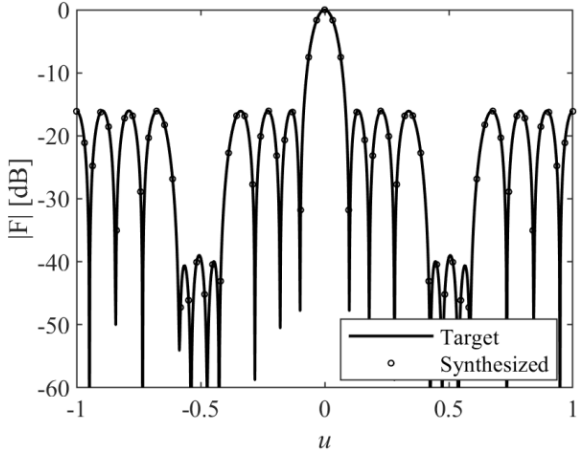

Fig. 5: The synthesized result of array with deep-null pattern.

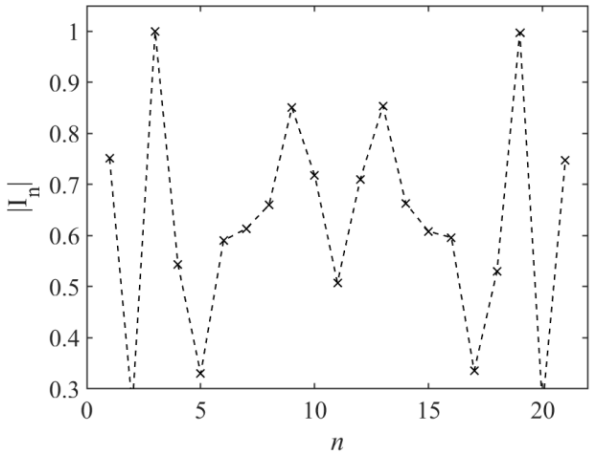

Fig. 6: The excitation coefficients of the third example.

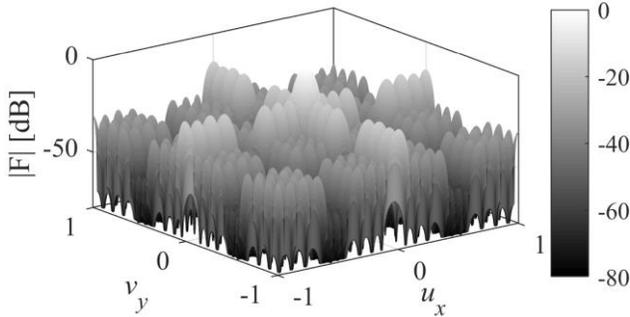

Fig. 7: The 3D synthesized result of planar array with deep-null pattern.

## D. Planar array with deep nulls

As mentioned in the previous section, the proposed method can be expanded to a planar array. So, the previous case is developed to a planar array with $21\times21$ elements and elements spacing $\lambda/2 \times \lambda/2$. The 3-dimensional array factor and 2-dimensional array weights are depicted in Figures (7) and (8), respectively. The accuracy of obtained results are acceptable.

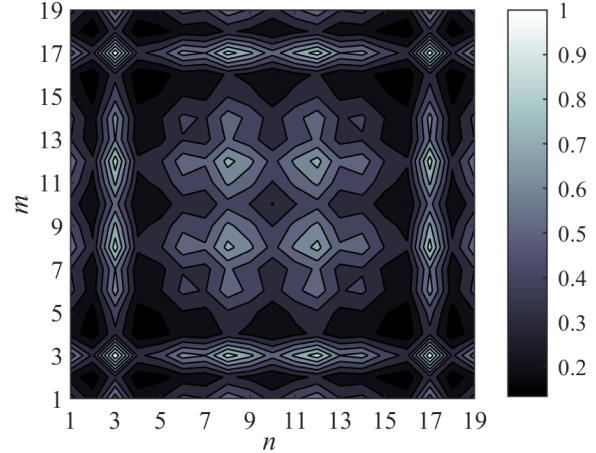

Fig. 8: The excitation coefficients of the fourth example.

## IV. CONCLUSION

Various iterative techniques, including Richardson method, Jacobi method, Gauss-Seidel method, Successive-over relaxation (SOR) method are investigated for synthesizing of the array factor of the linear and planar antenna arrays. Usually, least square method (LSM) is utilized for this purpose. But, LSM cannot be applied for any practical arrays. Because LSM is not an overall approach, and it should be combined with other techniques to determine both magnitude and phase of the excitation coefficients of each element of the under-studying array. Results show that this method is a good candidate for this purpose.